\documentclass{PoS}

\title{Compatibility of global NPDF analyses of neutrino DIS and charged-lepton DIS data}

\ShortTitle{Compatibility of global NPDF analyses of neutrino DIS and charged-lepton DIS data}

\author {
         K. Kovarik \\
         Karlsruhe, Institute of Technology, D-76128, Germany\\
        E-mail: \email{kovarik@particle.uni-karlsruhe.de}
}
\author {
        I. Schienbein  \\
         LPSC, Universit\'e Joseph Fourier Grenoble 1, CNRS/IN2P3, INPG, Grenoble, F-38026, France\\
        E-mail: \email{schien@lpsc.in2p3.fr}
}
\author {
        F.I. Olness \\
        Southern Methodist University, Dallas, TX75275, USA \\
        E-mail: \email{olness@physics.smu.edu}
}
\author{\speaker{J.Y. Yu}%
         \thanks{The work of J.Y. Yu was supported by the Deutsche Forschungsgemeinschaft(DFG) through grant Nr. YU 118/1-1.}\\
         LPSC, Universit\'e Joseph Fourier Grenoble 1, CNRS/IN2P3, INPG, Grenoble, F-38026, France \& Southern 
        Methodist University, Dallas, TX75275, USA\\
        E-mail: \email{yu@lpsc.in2p3.fr}}
\author {
        C. Keppel \\
       Jefferson Lab, Newport News, VA 23602, USA\& Hampton University, VA 23668, USA\\
        E-mail: \email{keppel@jlab.org}
}
\author {
        J.G. Morfin \\
         Fermilab, Batavia, IL 60510, USA\\
        E-mail: \email{morfin@fnal.gov}
}
\author {
        J.F. Owens \\
         Florida State University, Tallahasse, FL 32306-4350, USA\\
        E-mail: \email{owens@hep.fsu.edu}
}
\author {
         T. Stavreva\\
         LPSC, Universit\'e Joseph Fourier Grenoble 1, CNRS/IN2P3, INPG, Grenoble, F-38026, France\\
        E-mail: \email{stavreva@lpsc.in2p3.fr}
}

\abstract{The neutrino deep inelastic scattering (DIS) data is very interesting 
for global analyses of proton and nuclear parton distribution functions (PDFs) 
since they provide crucial information on the strange quark distribution in the proton 
and allow for a better flavor decompositon of the PDFs. 
In order to use neutrino DIS data in a global analysis of proton PDFs nuclear effects need to be understood. 
We study these effects with the help of nuclear PDFs 
extracted from global analyses of charged-lepton DIS, Drell-Yan and neutrino DIS data
at next-to-leading order in QCD. 
}

\FullConference{XXIst International Europhysics Conference on High Energy Physics\\
		 21-27 July 2011\\
		 Grenoble, Rhones Alpes France}

\begin{document}

\section{Introduction}
\noindent Parton distribution functions are of great importance in contemporary high energy physics.
Because they encode fundamental information on the structure of hadrons,
PDFs are needed for the computation of any high energy reaction 
(at HERA, RHIC, Tevatron, LHC, \dots) involving hadrons in the initial state.

Many groups perform and regularly update global analyses of PDFs for protons \cite{Ball:2009mk, 
Martin:2009iq, Nadolsky:2008zw,JimenezDelgado:2008hf} and nuclei \cite{Hirai:2007sx, Eskola:2009uj,deFlorian:2003qf}. 
Although not often emphasized, nuclear effects are present also in the analyses of proton PDFs since a number of experimental data is 
taken on nuclear targets. Most of the nuclear targets used in the proton analysis are made of light nuclei in 
which nuclear effects are expected to be small. 
However, the neutrino DIS data is taken on heavy nuclei such as iron and lead and is sensitive to 
the strange quark content of the proton. 
It should be noted that a good knowledge of the strange quark PDF has a significant influence on
the $W$- and $Z$ boson benchmark processes at LHC. 
Moreover, the neutrino experiments have been used to make precision tests of the Standard Model (SM) in the neutrino sector.
A prominent example is the extraction of the weak mixing angle $\sin\theta_W$ in a Paschos-Wolfenstein type analysis.

Historically, nuclear corrections based on information from the charged-lepton--nucleus ($\ell A$) DIS data have been applied 
to the neutrino--nucleus ($\nu A$) DIS data.
Furthermore, the {\em same correction factors} have been applied to several different observables in $\nu A$-DIS
($F_2$, $F_3$, cross section and dimuon production) and at different scales $Q^2$. 
Conversely, it is much more flexible to compute nuclear corrections in the Parton Model (PM) using nuclear PDFs
taking different observables and scales into account. 
For this reason, we perform global analyses of nuclear PDFs at next-to-leading order (NLO) in QCD 
in a framework closely related to the one used by the CTEQ collaboration in Ref.\ \cite{Pumplin:2002vw}. 
We determine several sets of nuclear PDFs from a) $\ell A$ DIS + Drell-Yan (DY) data and b) $\nu A$ DIS data
and c) the combined $\ell A$ DIS + DY + $\nu A$ DIS data.
We analyze and compare the resulting nuclear correction factors.   
\section{CTEQ Nuclear Parton distribution functions}
\noindent
The global nuclear PDF framework we use to analyze $\ell A$ DIS and DY and $\nu A$ DIS data 
was introduced in \cite{Schienbein:2009kk}. 
The parameterizations of the parton distributions in bound protons at the input scale of $Q_0=1.3\ {\rm GeV}$	
\begin{equation}
x\, f_{k}(x,Q_{0}) = c_{0}x^{c_{1}}(1-x)^{c_{2}}e^{c_{3}x}(1+e^{c_{4}}x)^{c_{5}}\,,\label{eq:input1}
\end{equation}
where $k=u_{v},d_{v},g,\bar{u}+\bar{d},s,\bar{s}$ and
\begin{equation}	
\bar{d}(x,Q_{0})/\bar{u}(x,Q_{0}) = c_{0}x^{c_{1}}(1-x)^{c_{2}}+(1+c_{3}x)(1-x)^{c_{4}}\,,\label{eq:input2} 
\end{equation}
are a generalization of the parton parameterizations in free protons used in the CTEQ proton analysis \cite{Pumplin:2002vw}. 
To account for a variety of nuclear targets, we introduce A-dependent fit parameters $c_k$
\begin{equation}
c_{k}\to c_{k}(A)\equiv c_{k,0}+c_{k,1}\left(1-A^{-c_{k,2}}\right),\ k=\{1,\ldots,5\}\,.\label{eq:Adep}
\end{equation}
The free proton PDFs are recovered in this framework in the limit $A\rightarrow 1$. 
From the input distributions, we can construct the PDFs for a general $(A,Z)$-nucleus 
\begin{equation}
f_{i}^{(A,Z)}(x,Q)=\frac{Z}{A}\ f_{i}^{p/A}(x,Q)+\frac{(A-Z)}{A}\ f_{i}^{n/A}(x,Q),\label{eq:pdf}
\end{equation}
where the distributions of a bound neutron, $f_{i}^{n/A}(x,Q)$, are related to those of a proton by isospin symmetry. 

We performed a global analysis of $\ell A$ DIS + DY data within this framework, 
determining the $A$-dependence of the 
parameters $c_k(A)$. In this analysis, we applied the same standard kinematic cuts $Q>2 \ {\rm GeV}$ and $W>3.5\ {\rm GeV}$ as in 
\cite{Pumplin:2002vw} leaving 708 data points 
and we obtained a fit with $\chi^{2}/{\rm dof}=0.946$  
with 32 free parameters (for further details see \cite{Schienbein:2009kk}). 

To extract the nuclear effects it is useful to define a nuclear correction factor $R$ 
which is the ratio of the observable ($\bf O$) using nuclear PDF over free PDF 
\begin{equation}
R[\bf O]=\frac{\bf O [nuc.\,\, PDF]}{\bf O [free\,\, PDF]}.
\label{eq:obs}
\end{equation}
We compared the nuclear correction factors $R$ for the structure function $[F_2^{\ell A}]$ in $\ell A$ DIS 
with the structure function $[F_2^{\nu A}]$ in $\nu A$ DIS.
As first observed in \cite{Schienbein:2007fs}, the nuclear correction factor 
$R[F_2^{\ell A}]$ does not describe the NuTeV $\nu Fe$ data well. 
This raises the question whether the nuclear corrections in $\ell A$ DIS and $\nu A$ DIS are different.

For definite conclusions, we 
set up a global analysis of the combined $\ell A$ DIS + DY + $\nu A$ DIS data  
where we included exclusively the neutrino DIS cross-section data coming from the 
NuTeV (iron) and Chorus (lead) experiments, respectively. 
Here we applied the same kinematic cuts as in the first analysis of the charged-lepton data 
leaving us with 3134 neutrino DIS cross-section data.
In a fit to {\em only} the neutrino data we obtained a $\chi^{2}/{\rm dof}$ of 1.33 
with 34 free parameters (for further details see \cite{Kovarik:2010uv}).

As expected, the nuclear correction factors using NPDFs
extracted from the $\nu A$ DIS data exhibit clear differences with the corresponding nuclear correction
factors using NPDFs extracted from $\ell A$ DIS + DY data 
which 
are especially marked at low and intermediate Bjorken $x$. 

Analyzing both data sets in a combined global analysis runs 
into the problem of an imbalance of the number of data points 
between the two data sets, since the $\nu A$ DIS data points would be dominant compared to the $\ell A$ DIS and DY data. 
Therefore, we introduced a weight parameter $w$ to combine the data sets
\begin{equation}
\chi^{2}=\sum_{l^{\pm}A\ {\rm data}}\chi_{i}^{2}\ +\!\!\sum_{\nu A\ {\rm data}}w\,\chi_{i}^{2}\ .\label{eq:chi2}
\end{equation}
In the case $w=0$, only the $\ell A$ DIS and DY data are included and $w=\infty$ uses only
the $\nu A$ DIS data. Varying the weight $w$, we tried to find a compromise fit which would describe both, $\ell A$ DIS + DY data and $\nu A$ DIS data, reasonably well.
We refer to Table II in \cite{Kovarik:2010uv} for the resulting $\chi^2$ of the compromise fits with weights $w=0,1/7,1/2,1,\infty$ 
and to Figure 1 in \cite{Kovarik:2010uv} for the corresponding nuclear correction factors. 
As one can see, with increasing weight $w$ the description of the $\ell A$ DIS data gets worse and the one of the $\nu A$ DIS data 
improves.

In order to judge quantitatively on how well the compromise fits describe the data we use the $\chi^2$ 
goodness-of-fit criterion introduced in \cite{Stump:2001gu,Martin:2009iq}. 
We consider a fit to be a good compromise if its $\chi^2$ for the $\ell A$ DIS + DY + $\nu A$ DIS data is 
within the 90\% confidence level 
of the fits to a) only $\ell A$ DIS + DY data and b) only $\nu A$ DIS data. 
We define the 90\% percentile $\xi_{90}$ used to define the 90\% confidence level, by
\begin{equation}\label{xi90}
	\int_0^{\xi_{90}}P(\chi^2,N)d\chi^2 = 0.90\,,
\end{equation}
where $N$ is the number of degrees of freedom and $P(\chi^2, N)$ is the probability distribution 
\begin{equation}\label{chi2dist}
	P(\chi^2,N) = \frac{(\chi^2)^{N/2-1}e^{-\chi^2/2}}{2^{N/2}\Gamma(N/2)}\,.
\end{equation}
We can assign a 90\% confidence level error band to the $\chi^2$ of the fits to the $\ell A$ DIS + DY and 
to the $\nu A$ DIS data 
\begin{equation}\label{lnuA90}
	\chi^2_{l^\pm A} = 638+ 45.6,\qquad \chi^2_{\nu A} = 4192+ 138.
\end{equation}
Comparing the results of the fits with different weights, listed in Table II in \cite{Kovarik:2010uv}, 
we conclude that there is no good compromise which
is compatible with both 90\% confidence level given in Eq.(\ref{lnuA90})
(see details in \cite{Kovarik:2010uv}).

\section{Conclusion}
\noindent After performing a thorough global NPDF analysis of the combined charged-lepton DIS, DY and neutrino DIS data, 
we find that nuclear correction factors in $\nu A$ DIS and $\ell A$ DIS are different and there is no good 
compromise fit to the combined $\ell A$ DIS, DY and $\nu A$ DIS data.
This has important consequences for global analyses of proton and NPDF, for models explaining 
the nuclear effects, and precision observables in the neutrino sector. 
The nCTEQ nuclear PDFs described here as well as the ones obtained in \cite{Stavreva:2011} are available at \cite{ncteq:2011}.

\end{document}